\documentclass{nature}
\usepackage{graphicx}
\makeatletter
\let\saved@includegraphics\includegraphics
\AtBeginDocument{\let\includegraphics\saved@includegraphics}
\usepackage[font=footnotesize]{caption}
\renewenvironment*{figure}{\@float{figure}}{\end@float}
\makeatother
\usepackage{hyperref}
\usepackage{color}
\definecolor{dark-gray}{gray}{0.15}
\title{Generation and annihilation time of magnetic droplet solitons}

\author{Jinting Hang$^{+1}$, Christian Hahn$^{+1}$, Nahuel Statuto$^{2,3}$, Ferran Maci{\`a}$^{2,3}$ and Andrew D. Kent$^{*1}$}

\begin{document}

\maketitle
\vspace{-0.5 cm}
\begin{affiliations}
\item{Center for Quantum Phenomena, Department of Physics, New York University, New York, New York 10003 USA}
\item{Department of Condensed Matter Physics, University of Barcelona, 08028 Barcelona, Spain}
\item{Institut de Ci\`encia de Materials de Barcelona (ICMAB-CSIC), Campus UAB, 08193 Bellaterra, Spain}\\
$^{+}$These authors contributed equally to this work. \\
$^*$Corresponding author: 
\href{mailto:andy.kent@nyu.edu}{\color{blue} andy.kent@nyu.edu}
\end{affiliations}

\begin{abstract}
Magnetic droplet solitons were first predicted to occur in materials with uniaxial magnetic anisotropy due to a long-range attractive interaction between elementary magnetic excitations, magnons. A non-equilibrium magnon population provided by a spin-polarized current in nanocontacts enables their creation and there is now clear experimental evidence for their formation, including direct images obtained with scanning x-ray transmission microscopy. Interest in magnetic droplets is associated with their unique magnetic dynamics that can lead to new types of high frequency nanometer scale oscillators of interest for information processing, including in neuromorphic computing. However, there are no direct measurements of the time required to nucleate droplet solitons or their lifetime---experiments to date only probe their steady-state characteristics, their response to dc spin currents. Here we determine the timescales for droplet annihilation and generation using current pulses. Annihilation occurs in a few nanoseconds while generation can take several nanoseconds to a microsecond depending on the pulse amplitude. Micromagnetic simulations show that there is an incubation time for droplet generation that depends sensitively on the initial magnetic state of the nanocontact. An understanding of these processes is essential to utilizing the unique characteristics of magnetic droplet solitons oscillators, including their high frequency, tunable and hysteretic response.

 \end{abstract}

Magnetic droplet solitons~\cite{Ivanov1977} form in spin transfer torque oscillators~\cite{Hoefer2010,Mohseni2013,Macia2014,Kiselev2003,Rippard2004,Muduli2010} that consist of nanometer scale contacts to thin films with perpendicular magnetic anisotropy~\cite{Mohseni2013,Macia2014,Backes2015,Chung2016,Chung2017}, as predicted by theory~\cite{Hoefer2010}.  They have been studied in thin films with two ferromagnetic layers, a free layer and a fixed spin-polarizing layer, separated by a thin metallic layer. Droplets consist of a partially reversed coherently precessing magnetization in and near the nanocontact region. Through the giant magnetoresistance effect, droplet formation leads to a step increase in nanocontact resistance as well as a sharp peak in the high-frequency spectra at the spin-precession frequency when overcoming a threshold current~\cite{Mohseni2013,Macia2014}. There have been a number of proposed applications of magnetic droplet solitons that rely on their unique magnetic dynamics~\cite{Bonetti2014}. First, their output signal can be larger because of their large angle spin excitations giving a larger magnetoresistive response \cite{Bonetti2012,Chen2016}. [While magnetic droplet solitons have been studied thus far in metallic multilayers, forming droplets in a magnetic tunnel structure would provide a larger output signal because of their larger magnetoesistive response.]  Second, their response is highly tunable with current and hysteretic~\cite{Macia2014}. Third, their oscillation frequencies can be one to two orders of magnitude larger than that of magnetic vortex oscillators~\cite{Pribiag2007}, recently used in neuromorphic computing demonstrations~\cite{Torrejon2017,Romera2017}, thus enabling higher speed device operation. Applications require knowledge of the time required to nucleate droplet solitons and their lifetime as well as an understanding of the nucleation and annihilation processes. Thus far experiments only probe the steady-state characteristics of droplet solitons, i.e. their properties when the current has been on or off for long periods compared to the time scale of their intrinsic dynamics.

In this article we report measurement of the current pulse times for droplet generation and annihilation. In these studies we use the fact that droplet solitons exhibit magnetic bistability; that there are large ranges of applied current and field at which both droplet and non-droplet states are possible \cite{Macia2014}. The nanocontact state thus depends on its prior history. By biasing a nanocontact in a bistable condition and applying current pulses we determine the pulse amplitude and duration needed to generate and annihilate droplet solitons. 

The experiments were performed on nanocontacts to metallic multilayers that consist of a perpendicularly magnetized free layer and easy-plane fixed layer (Fig.~1a). These layers are separated by a non-magnetic spacer layer that decouples them magnetically while enabling a flow of spin-polarized current. (See the Methods Section for the layer compositions.) All measurements reported here were done at room temperature.  

A step increase/decrease in dc-resistance is associated with the creation/annihilation of the droplet state, which can be used for determining the critical current needed to drive the droplet dynamics. The step in resistance corresponds to an abrupt change in the high-frequency spectra of the nanocontact, consistent with droplet creation (a step decrease in the noise frequency) and annihilation (a step increase in the noise frequency), as reported in Refs.~\cite{Mohseni2013,Macia2014} (see Supplementary Section II for nanocontact spectral data.) For a certain field range (see Supplementary Section I and Ref.~\cite{Macia2014}), one finds a current zone where the droplet is bistable, i.e. both droplet and non-droplet state are possible for the same current.
For a fixed applied magnetic field in the bistable zone, two critical currents are observed: $I_{c1}$ below which the droplet state does not form and $I_{c2}$ ($> I_{c1})$ above which there is only a droplet state. Between the two critical currents a hysteretic response is observed. In Fig.~1b we show the nanocontact resistance at a fixed applied field of 0.7~T while ramping the current up (red curve) and down (blue curve).  Here we find $I_{c1}= 12.3 \pm 0.07$ mA and $I_{c2} = 14.2 \pm 0.14$~mA. The uncertainly is the standard deviation in the currents $I_{c1}$ and $I_{c2}$ determined by $50$ repeated I-V measurements. The distributions of $I_{c1}$ and $I_{c2}$ are shown as histograms that are overlaid on Fig.~1b. We note that the standard deviation of the generation current is twice that of the annihilation current, showing there is a greater stochasticity in the generation process.

To study droplet generation and annihilation we bias the nanocontact with a current in the hysteretic region and apply pulses to momentarily change the current. For example, to study droplet annihilation we initialize the nanocontact in the droplet state by ramping the current up to $ I  > I_{c2}$  and then reduce the current to a range $ I_{c1} < I  < I_{c2}$. At a current within the hysteretic region (indicated by the dashed line in panels b-d of Fig.~1), we apply a negative polarity pulse to momentarily reduce the current below $I_{c1}$. There is a step down in nanocontact resistance if the droplet has been annihilated, as seen in Fig.~1c. To study droplet generation, we start in a non-droplet state, again in the hysteretic region $I_{c1} < I  < I_{c2}$ (indicated by the dashed line in Fig.~1d), and apply a positive polarity current pulse to increase the current above $I_{c2}$. Droplet generation causes a step increase in the nanocontact resistance as seen in Fig.~1d. We repeat these pulses multiple times to determine the generation and annihilation probabilities as a function of pulse amplitude and duration.

\begin{figure}
   \centering
 \includegraphics[width=0.5\textwidth]{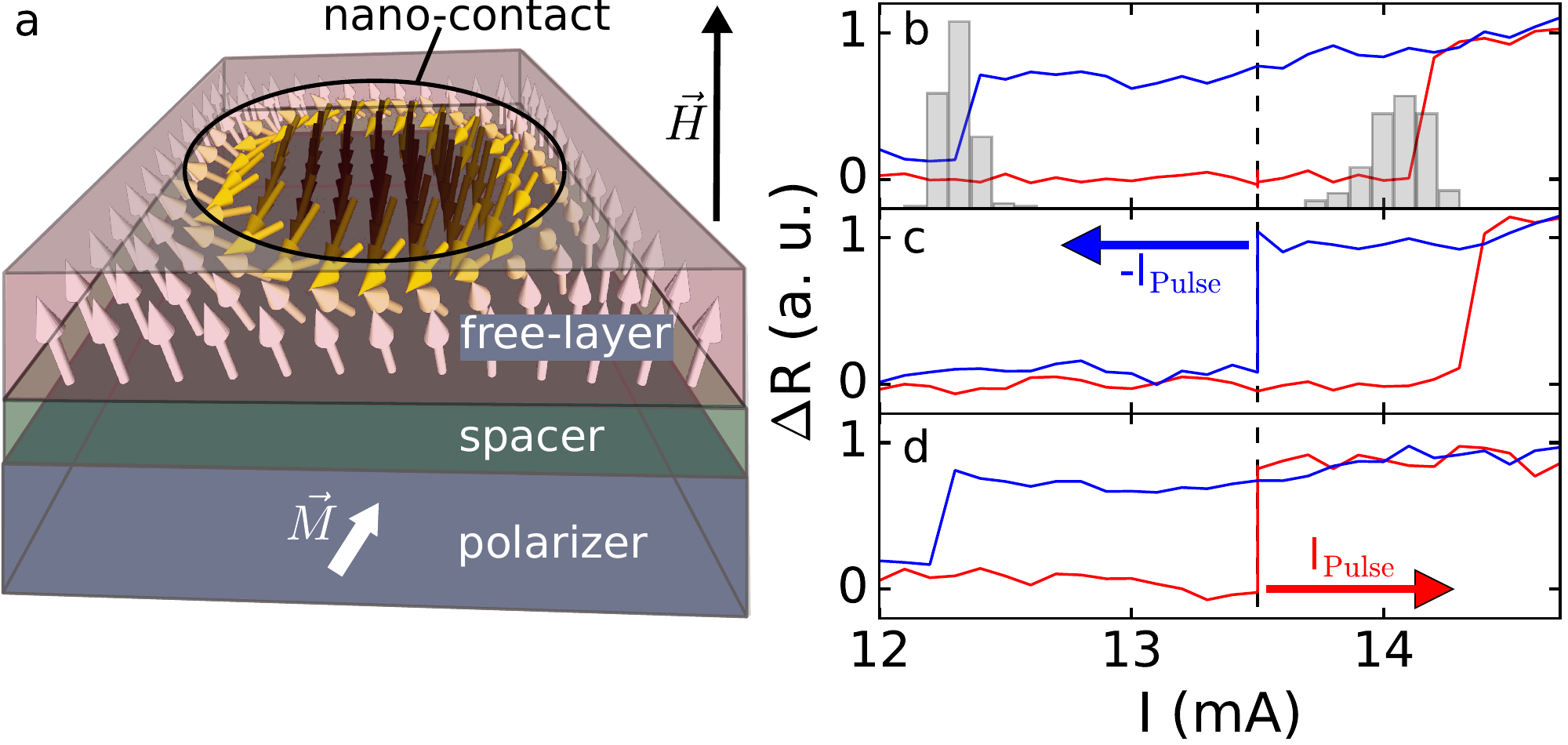}
 \linespread{1}
   \caption{{\bf a}, Schematic of a multilayer with a magnetic droplet. Electrons are injected through a nanocontact to the free layer and move from the free layer (red) to the polarizer (blue) for positive current polarity. An external field is applied perpendicularly to the film plane, partly canting the polarizer magnetization. {\bf b}, Resistance vs. current at a fixed field of 0.7 T after subtracting a background caused by Joule heating. The overlaid histogram indicates the distribution of generation and annihilation currents. {\bf c}, The nanocontact is biased at 13.5 mA in the higher resistance state (dashed line). A negative current pulse annihilates the droplet, as seen by the step decrease in nanocontact resistance. {\bf d}, Starting in the non-droplet state a positive current pulse can generate the droplet, as seen by the step increase in nanocontact resistance.}
\label{Fig:Device}
\end{figure}

The droplet annihilation probability as a function of pulse duration for different pulse amplitudes is shown in Fig.~2.
We indicate the standard error of the mean  as error bars, SEM$= \surd({\sum}_{i=0}^{n} (x_i-\mu)^2/((n-1)n))$, with $x_i$ as the individual trial result, ${\mu}$ being the mean value and $n$ the number of trials, $n=50$ for the results in Figs.~2 \& ~3.
In Fig.~2, we observe a monotonic increase of the annihilation probability as a function of pulse duration. Varying the applied pulse amplitude from 2.24~mA to 3.15~mA we also see an increase in the probability with increasing pulse amplitude, reaching 100~\% annihilation probability at 2~ns for 3.15~mA.

\begin{figure}
  \centering
    \includegraphics[width=0.5\textwidth]{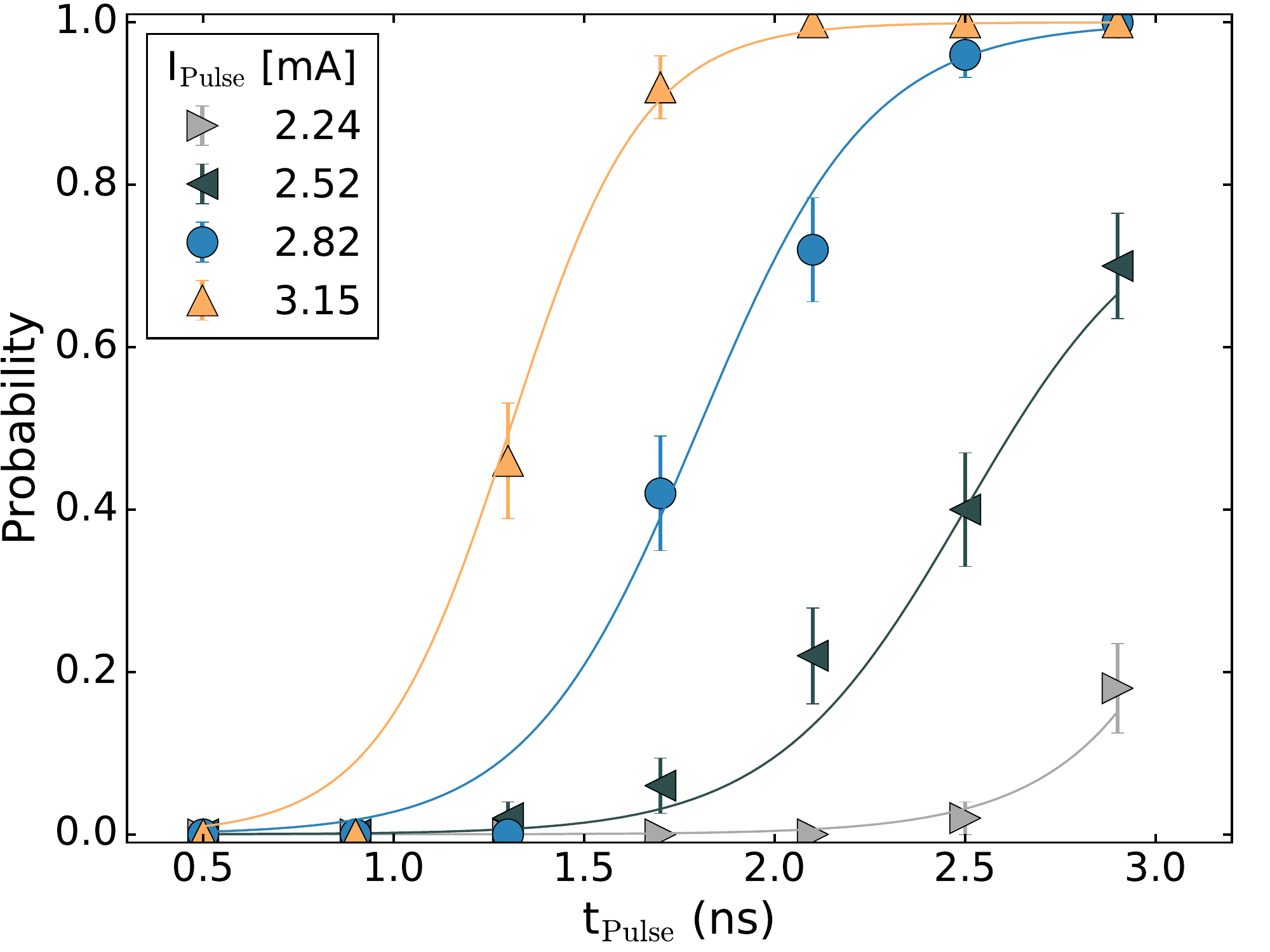}
     \linespread{1}     
     \caption{The annihilation probability versus pulse duration at different pulse amplitudes. Pulses were added to a dc current of 13.5mA and a 0.7~T field was applied. The lines are guides to the eye.}
\label{Fig:Annihaltion}
\end{figure}

We then proceed to study the times needed to generate the droplet by initializing the nanocontact in the non-droplet state and applying positive current pulses.  Again, we start with a bias current of 13.5 mA and apply varying pulse durations and amplitudes that briefly increases the current above the threshold current $I_{c2}$.
The probability for droplet generation is shown as a function of pulse duration for different pulse amplitudes in Fig.~3. Similarly to the annihilation probability we observe a  monotonic increase with the pulse duration. The pulse duration is plotted on a logarithmic scale to encompass the range of generation times observed. For the highest pulse amplitude applied in the annihilation experiment (3.15~mA), we obtain 50\% generation probability at a pulse duration of 70~ns. We achieve 50\% creation probability for a 3.93~mA amplitude and 10~ns duration pulse. We note that we have also examined samples with 150~nm diameter nanocontacts. These nanocontacts show similar behavior, exhibiting a monotonic increase of generation and  annihilation probability with pulse duration; they display an even stronger asymmetry between the generation and annihilation pulse durations, necessitating yet one order of magnitude longer pulses for droplet generation than shown in Fig.~3 yet still short (ns) pulses for annihilation. (Data on a 150~nm diameter nanocontact is shown in Supplementary Section III.)

\begin{figure}
  \centering
    \includegraphics[width=0.5\textwidth]{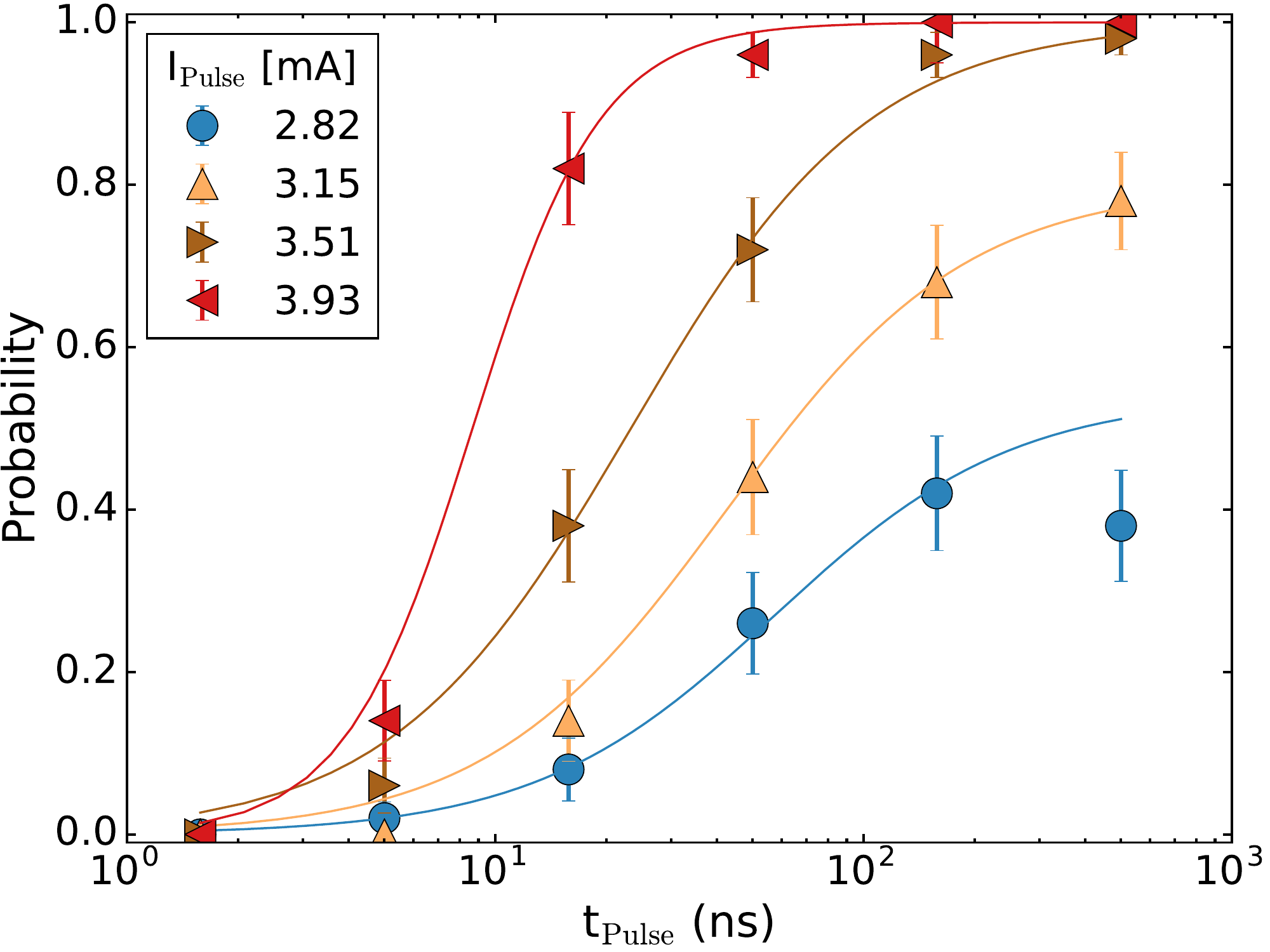}
     \linespread{1}
     \caption{The generation probability versus pulse duration for different pulse amplitudes. The dc current was fixed at 13.5mA
      and a 0.7~T field was applied. The lines are guides to the eye.}
\label{Fig:Generation}
\end{figure}

These results show that the time needed to enter the droplet state can be longer than for exiting it; for the same current pulse amplitude it can take orders of magnitude longer to create the droplet than to annihilate it. This indicates a different mechanism involved in annihilation and generation processes. To better understand this asymmetry we modeled the generation and the annihilation processes of magnetic droplets. We consider a circular nanocontact to a ferromagnetic thin film with perpendicular magnetic anisotropy with parameters taken from experiment. (Parameters are given in the Methods Section.) The inset in Fig.~4b, left panel, shows the average z-component of magnetization in the nanocontact $m_z$ as a function of the applied current. Clear hysteresis is seen, like in the experiment; higher current is required to generate the droplet with increasing current, while droplet annihilation occurs at lower current when decreasing the current. 

To study droplet annihilation and generation we start at a dc current in the hysteretic zone, where both droplet and non-droplet states are possible. As in the experiment, the current is momentarily reduced to study droplet annihilation and increased to study droplet creation.
Fig.~4b shows the time evolution of the magnetization of the nanocontact in the annihilation process, left panel, and for the generation process, right panel. Fig.~4c shows the applied current as a function of time. In the annihilation process a negative polarity pulse of $-1.7$ mA is applied to a dc current of $13.8$ mA. The pulse duration needed to annihilate is $1.3$ ns. On the other hand, in the generation process, starting in a non-droplet state, a (positive polarity) pulse of $1.7$ mA is applied. Under these conditions it takes $21$ ns to create a droplet. After $21$ ns the current is reduced to the initial current ($13.8$ mA) and this current sustains the droplet. We thus see that for the same pulse amplitude the generation and annihilation processes occur on different time scales. The generation process has a waiting time and after that time the droplet forms rapidly, in a transition time (20 \% to 80\% of the initial to final $m_z$) {\em of just nanoseconds}. The images in Fig.~4a show the magnetization at different times in the generation and the annihilation processes.  We note that the final state of the generation process and the initial state in annihilation process are not the same, as after some time the droplet moves in the nanocontact and the spins at the boundary of the droplet dephase \cite{Mohseni2013} (see Supplementary Section IV). 

It is important to note that a long droplet generation time is not fundamental; the droplet generation time can be reduced and made comparable to the annihilation time by a number of means. For example, larger current pulse amplitudes lead to 10 ns generation times, as seen in Fig.~3 for $I_\mathrm{pulse}=3.93$~mA. Further the incubation time is a very sensitive function of the initial magnetization state of free layer. An initial state with the magnetization tilted with respect to anisotropy axis by just a few degrees leads to nanosecond generation times, as discussed in the Supplementary Sections V and VI. This can be achieved by applying an in-plane field pulse (see Supplementary Section VI). Finally, while in our experiments the nanocontact is initially biased with a dc current, droplets can also be generated by applying pulses starting with a no bias current (see Supplementary Section VI). A bias current is applied in our experiments for a practical reason---it enables a voltage readout of the nanocontact state after a pulse.
\begin{figure}
  \centering
    \includegraphics[width=0.8\textwidth]{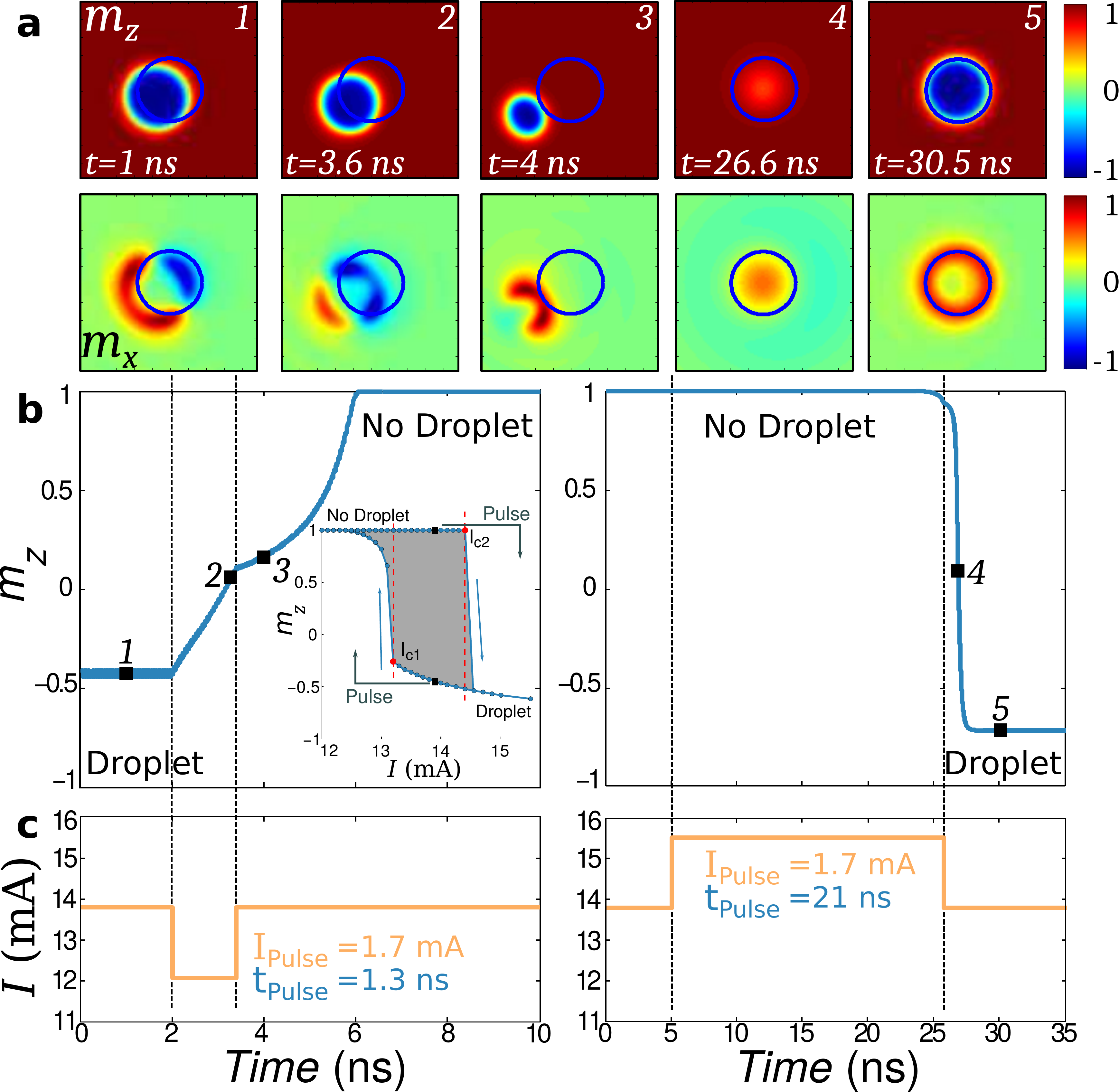}
     \linespread{1}
     \caption{Creation and annihilation process. {\bf a}, Images of the magnetization at times in the simulation. Images correspond to a $290\times 290$ nm$^2$ field of view. The blue circle shows the boundary of the nanocontact.
{\bf b}, Time evolution of the nanocontact magnetization for the annihilation and the creation processes, left and right images, respectively. The black squares correspond to the times shown in the images.  {\bf c}, Current applied as a function of time for the annihilation and generation processes. The vertical dashed black lines shows the time where the pulse was applied. The inset of {\bf b}, left hand panel, shows the average magnetization of the nanocontact as a function of the current and the resulting hysteresis: $I_{c1}=13.1$ mA and $I_{c2}=14.5$ mA and between these currents both droplet and non-droplet states are stable.}    
\label{Fig:panel}
\end{figure}

The duration of current pulses needed for generation and annihilation are thus different. It is interesting to compare the droplet annihilation time to the magnon relaxation time $\tau_\mathrm{m}=1/(2 \pi \alpha f)$, where $\alpha$ is the damping constant and $f$ is the spin-precession frequency.  For $\alpha = 0.03$ and a spin-precession frequency of 20 GHz, the magnon relaxation time is 0.2 ns, about an order of magnitude smaller than the observed droplet annihilation time. In micromagnetic simulations the droplet drifts across the boundary of the nanocontact, as seen in Fig.~4a, and dissipates within in a few nanoseconds after having left the region experiencing a spin-transfer torque. In contrast to droplet annihilation, where the magnetization relaxes to equilibrium, the generation process can be viewed as a transition between magnetic states. In this case, similar to the switching of a macrospin between two stable magnetic states by a spin-transfer torque, there is an incubation time associated with building up spin precession angle starting from a small initial angle~\cite{Sun2000,Devolder2008}. This can also be viewed as a time required to generate a sufficient number of magnons to form a droplet solition. Once initiated (i.e. after the waiting time) the formation process is fast, comparable to the times needed for droplet annihilation (several ns and less). We find that the incubation time becomes smaller with increasing initial magnetization angle and generation pulse amplitude (see Supplementary Section IV). 

In summary, we report measurement of the timescales for droplet generation and annihilation, which has led to an improved understanding of these processes. This understanding allows new means to control and modify these processes. For example, the droplet incubation time can be greatly reduced by increasing the pulse amplitude or creating an initial tilt of the magnetization.  We also note that the measured annihilation time underestimates the actual droplet lifetime, as it is possible for the droplet to move away from the nanocontact and still exist after application of the pulse. Finite temperature and noise processes must also play a role in generation and annihilation processes. These require additional analysis and micromagnetic simulations that we have not considered in this work. It will be interesting to study dynamical skyrmions~\cite{Chen2016,Statuto2018}, which are expected to be longer lived excitations, as well as to examine the effect of temperature on the stability and formation of droplet solitons.

\begin{methods}
The multilayers are composed of 10 Ni$_{80}$Fe$_{20}$ $|$10 Cu$|$[0.2 Co$|$0.6 Ni]$\times$6 (numbers before the elements are the layer thicknesses in nm) deposited on a 50 nm Cu bottom electrode on an oxidized Si wafer. The multilayer was encapsulated in a 50 nm thick SiO$_2$ dielectric layer. Permalloy (Ni$_{80}$Fe$_{20}$) is the fixed layer and CoNi is the free layer and has perpendicular magnetic anisotropy. The Cu layer between the fixed and free layer is thinner than its spin-diffusion length enabling effective spin-transport between the layers. The effective perpendicular anisotropy field $M_\mathrm{eff}$ of the CoNi layer was determined using ferromagnetic resonance spectroscopy with the applied field perpendicular to the film plane, $\mu_0 M_\mathrm{eff}  = \mu_0(H_k - M_s )=  0.25$ T, where $\mu_0$ is the permeability of free space, $H_k$ perpendicular anisotropy field, and $M_s$ is the saturation magnetization \cite{Macia2012}.

E-beam lithography was used to define the nanocontact, a $80-150$ nm diameter aperture in resist. The resist pattern was transferred into the SiO$_2$ dielectric capping layer using reactive ion etching. The resulting device structure is show in Fig. 1(a), where the point-contact is indicated as a non-shaded area on top of the free layer. Electrons flow from the free layer to the polarizing layer for positive current polarity ($I>0$). The field is applied perpendicular to the film plane to cant the magnetization of the polarizer partly out of the plane. The free layer magnetization dynamics is illustrated by a grid of arrows. In the area of current flow, the out of plane magnetized free layer has nearly reversed its magnetization, as it is the case when $I > I_{c2}$. We use the high frequency port of a bias-T to couple short pulses into a nanocontact and contact the sample using a ground-signal-ground probe.

The parameters for the micromagnetic simulations were: saturation magnetization, $M_s=5 \cdot 10^5$ A/m, damping constant $\alpha=0.03$, uniaxial anisotropy constant $K_u=2 \cdot 10^5$ J/m$^3$, exchange stiffness $A=10^{-12}$ J/m and a nanocontact diameter of $100$ nm. We assumed an electrical current with a spin polarization $p=0.21$ passing through the nanocontact and an applied field of $0.7$ T out of the film plane.  We performed micromagnetic simulations using the open-source MuMax$^3$ code~\cite{mumax3} with a graphics card with 2048 processing cores. We considered the effects of Oersted fields but not interfacial Dzyaloshinskii-Moriya interactions (DMI) or finite temperature. The simulation of $m_z$ as a function of the applied current in the inset of Fig.~4b, left panel, was performed with $80$ ns of simulation time for each applied current. Pulse current simulation start with a dc current of $I=13.8$ mA, which is in the hysteretic zone. The full code is available in Supplemental Section V.

\end{methods}

\section*{Acknowledgements}
F.M. acknowledges financial support from the Ram\'on y Cajal program through Grant No. RYC-2014-16515 and from MINECO through the Severo Ochoa Program for Centers of Excellence in R\&D (Grant No. SEV-2015-0496). N.S. acknowledges funding from SURDEC through the research training grant FI-DGR. Research at UB is partially supported through project MAT2015-69144-P (MINECO/FEDER, UE). Research at NYU was supported by Grant No. NSF-DMR-1610416.

\section*{References}

\begin{addendum}
\item F.M. acknowledges financial support from the Ram\'on y Cajal program through Grant No. RYC-2014-16515 and from MINECO through the Severo Ochoa Program for Centers of Excellence in R\&D (Grant No. SEV-2015-0496). N.S. acknowledges funding from SURDEC through the research training grant FI-DGR. Research at UB is partially supported through project MAT2015-69144-P (MINECO/FEDER, UE). Research at NYU was supported by Grant No. NSF-DMR-1610416.
\vspace{-0.5 cm}
 \item[Contributions]
J. Hang and C. Hahn contributed equally to this work. CH and ADK conceived of the experiment. JH and CH conducted the experiments and analyzed the experimental results. NS and FM modeled the experiments. All the authors discussed the results and wrote the manuscript.

\end{addendum}


\end{document}